\documentclass[epj,final]{svjour}

\usepackage{amssymb}
\usepackage{amsmath}
\usepackage{array}
\usepackage{graphicx}

\def\c{\cdot}

\def\adj{^\dagger}
\renewcommand\vec[1]{{\mathbf #1}}
\def\nul{_{\raisebox{1ex}{\scriptsize{(0)}}}}
\def\kb{k_{\mathrm{B}}}
\def\d{\mathrm{d}}
\def\conj{^\star}
\def\lb{\lambda_{\mathrm{B}}}
\def\Ordo{\mathcal{O}}
\def\gn{{g_{0}}}
\def\U{\mathrm{U}}
\def\ph{\varphi}
\allowdisplaybreaks

\begin{document}
\title{Renormalization group approach to the spin-1 Bose gas}
\author{Gergely Szirmai}
\institute{Research Group for Statistical Physics of the Hungarian
  Academy of Sciences, P{\'a}zm{\'a}ny P{\'e}ter S{\'e}t{\'a}ny 1/A,
  Budapest, H-1117}

\date\today

\abstract{
  A field theoretical renormalization group approach at two loop level
  is applied to the homogeneous spin-1 Bose gas in order to
  investigate the order of the phase transition. The beta function of
  the system with $d=4-\epsilon$ dimensions is determined up to the
  third power of the coupling constants and the system's free energy
  on the border of the classical stability is given in next to leading
  order. It is found that the phase transition of the interacting
  spin-1 Bose gases with weak spin-dependent coupling constant values
  is of first order.}



\maketitle

\section{Introduction}
\label{sec:intro}

Bose--Einstein condensation (BEC) of dilute, interacting, scalar
atomic gases is widely believed to be a continuous phase transition.
Mean-field theory results based on perturbation theory are however
contradicting.  The simplest approximations call for a continuous
phase transition \cite{SzK}, while according to the more sophisticated
Hartree--Fock (Popov) approximation BEC is of first order
\cite{Popov2,SG,FRSzG}. The contradicting results are understood to be
the consequence of critical phenomena. Namely, close to the critical
point, the different Feynman diagrams develop to the same magnitude,
and a perturbative treatment becomes inaccurate. The different
renormalization group calculations are indicating that BEC is a
continuous phase transition falling into the same universality class
as the $\mathrm{O}(2)$ model of field theory
\cite{BS2,AS1,MA1,Andersen1}.

In experiments made with dilute gases of alkali atoms in optical traps
\cite{SKea,Sea1,Sea2,Miesea,Stamper-Kurn2001a,BSC} the particles have
internal spin degrees of freedom. Such systems at low temperatures can
be modeled by Hamiltonians with multiple coupling constant
interactions \cite{Ho2,OM,HYin} in the s-wave scattering limit. For
such a situation, with multiple coupling constants, (and in the
homogeneous case) the order of the transition is not necessarily of
continuous type, as e.g. in the case of the field theoretical $\phi^4$
model with cubic anisotropy, where fluctuations can induce the
transition to be of first order \cite{Wallace1,KW1}.

In this paper we treat the problem of the homogeneous, spin-1 Bose
gas, where two coupling constants arise naturally in the low energy
limit. In the absence of an external magnetic field depending on the
magnitude of these two parameters two possible Bose--Einstein
condensed phases exist, namely, the ferromagnetic phase, in which the
system favors a macroscopic magnetization and the polar phase, in
which the system prefers no magnetization. The gas condensing to the
ferromagnetic phase is called the ferromagnetic gas, while the other
is the polar gas (see Eqs. \eqref{eqs:projmat} and the discussion
below). We assume zero external magnetic field. For such a system
mean-field theory results are also contradicting. In the Hartree--Fock
approximation \cite{SzSzK} the phase transition of both the polar and
ferromagnetic gases is of first order. The jump in parameters is the
function of the bigger coupling constant, which is responsible for non
spin-flip scatterings.  However this strongly first order type of
transition is considered to be an artifact of the Hartree--Fock
approximation such as in the case of the scalar Bose gas. On the other
hand the Hartree approximation \cite{SzSz2}, which is in a way a
simpler mean field approximation yields a continuous Bose condensation
in the polar case, while a first order one in the case of a
ferromagnetic Bose gas.  In the latter case the jump is a function of
the smaller coupling constant, responsible for spin flip scatterings.
(The ratio of the two coupling constants is typically in the order of
$10^{-2}$.) It is important to note that the Hartree approximation is
supplying a continuous BEC in the case of scalar Bose gases
\cite{SzK}, and it is related to the leading order approximation of
the $1/N$ expansion of the $\mathrm{O}(N)$ symmetric model in field
theory \cite{Amit1,Zinn-Justin1}. Because of the above ambiguities a
renormalization group approach is worked out to study the order of the
phase transition of the homogeneous, spin-1 Bose gas.  The formulation
is based on the assumption that the phase transition is continuous. In
this case the universal quantities (to leading order in the coupling
constants) and the infrared (IR) behavior of the system can be
calculated from a classical field theory obtained by restricting the
quantum fields to the zero Matsubara frequency sector
\cite{Landsman1,Andersen1}. The main conclusion of the paper is that
this assumption leads to contradiction which indirectly proofs that
the transition is of first order.

The paper is organized as follows. In Section \ref{sec:cft} the model
of the homogeneous spin-1 Bose gas is mapped to the corresponding
classical field theory by neglecting the nonzero Matsubara frequency
components of the quantum fields. In Section \ref{sec:renorm} the
renormalization program with minimal subtraction and dimensional
regularization is worked out up to the order of two loops for the
classical field theory. In section \ref{sec:crprop} the beta function
of the model is given up to the third power of the coupling constants
and the renormalized free energy is derived up to one loop level. The
critical properties of the system are discussed also in this section.
Some summary is left to Section \ref{sec:sum}.

\section{Classical field theory}
\label{sec:cft}

The effective Hamiltonian of the low temperature, homogeneous, spin-1
Bose gas can be written as
\begin{multline}
  \label{eq:ham}
  \mathcal{H}=\int\d^3x\Big[\frac{\hslash^2}{2M}\nabla\Psi\adj_r(x)
  \nabla\Psi_r(x)-\mu\Psi\adj_r(x)\Psi_r(x)\Big]\\+\frac{1}{2}\int\d^3x\;\d^3x'
  V^{rs}_{r's'}(x-x')
  \Psi\adj_r(x)\Psi_{r'}\adj(x')\Psi_{s'}(x')\Psi_s(x),
\end{multline}
with $M$ the mass of an atom and $\mu$ the chemical potential. The
bosonic field operator $\Psi\adj_r(x)$ creates an atom at position $x$
with spin projection $r\in\{+1,0,-1\}$, and the operator $\Psi_r(x)$
destroys it. Automatic summation over repeated spin indices is
implicitly assumed throughout the paper. The two-particle interaction
is modeled by s-wave scattering, with the interaction potential
\cite{Ho2,OM}
\begin{equation}
  \label{eq:pot1}
  V^{rs}_{r's'}(x-x')=\delta^{(3)}(x-x')\frac{4\pi\hslash^2}{M}\bigg[a_0\Big(\mathcal{P}_0
  \Big)^{rs}_{r's'}+a_2\Big(\mathcal{P}_2\Big)^{rs}_{r's'}\bigg],
\end{equation}
where $a_0$ and $a_2$ are the s-wave scattering lengths in the total
hyperfine channel $F=0$ and $F=2$, respectively. The matrices
$\mathcal{P}_0$ and $\mathcal{P}_2$ are the projection operators in
the 9-dimensional tensor product space of the spin variables
projecting to the subspaces of total hyperfine spin $0$ and $2$,
respectively. The matrix $\mathcal{P}_1$, projecting to the total
hyperfine spin-1 subspace is omitted from Eq. \eqref{eq:pot1}, since
it is antisymmetric in its indices and therefore does not appear in
the Hamiltonian \eqref{eq:ham}. The projection matrices can be
obtained from the following linear equations:
\begin{subequations}
  \label{eqs:projeqs}
  \begin{align}
    \mathcal{P}_0+\mathcal{P}_1+\mathcal{P}_2&=1,\\
    -2\mathcal{P}_0-\mathcal{P}_1+\mathcal{P}_2&=\vec{S}_1\c
    \vec{S}_2,\\
    4\mathcal{P}_0+\mathcal{P}_1+\mathcal{P}_2&=
    (\vec{S}_1\c\vec{S}_2)^2,
  \end{align}
\end{subequations}
where $\vec{S}_i$ $(i=1,2)$ is the spin operator of the i.th atom.
Using the usual spin-1 operators in the basis of $S_z$ eigenvectors
one can get from Eqs.  \eqref{eqs:projeqs} the projection matrices
$\mathcal{P}_0$, $\mathcal{P}_1$ and $\mathcal{P}_2$. Only expressing
the needed two, $\mathcal{P}_0$ and $\mathcal{P}_2$ are given by:
\begin{subequations}
  \label{eqs:projmat}
  \begin{align}
    \Big(\mathcal{P}_0\Big)_{RS}&=\frac{1}{3}\left[
      \begin{array}{c c c c c c c c c}
        0&0&0&0&0&0&0&0&0\\
        0&0&0&0&0&0&0&0&0\\
        0&0&1&0&-1&0&1&0&0\\
        0&0&0&0&0&0&0&0&0\\
        0&0&-1&0&1&0&-1&0&0\\
        0&0&0&0&0&0&0&0&0\\
        0&0&1&0&-1&0&1&0&0\\
        0&0&0&0&0&0&0&0&0\\
        0&0&0&0&0&0&0&0&0
      \end{array}
    \right],\\
    \Big(\mathcal{P}_2\Big)_{RS}&=\frac{1}{6}\left[
      \begin{array}{c c c c c c c c c}
        6&0&0&0&0&0&0&0&0\\
        0&3&0&3&0&0&0&0&0\\
        0&0&1&0&2&0&1&0&0\\
        0&3&0&3&0&0&0&0&0\\
        0&0&2&0&4&0&2&0&0\\
        0&0&0&0&0&3&0&3&0\\
        0&0&1&0&2&0&1&0&0\\
        0&0&0&0&0&3&0&3&0\\
        0&0&0&0&0&0&0&0&6
      \end{array}
    \right],
  \end{align}
\end{subequations}
with $R=5-3r-r'$ and $S=5-3s-s'$.

As shown by Ho \cite{Ho2} and Ohmi and Machida \cite{OM}, if $a_2>a_0$
the low temperature, equilibrium phase has a macroscopic wave function
with zero net magnetization, called as the polar phase in analogy to
the $^3\mathrm{He}$ case, while if $a_0>a_2$ the low temperature phase
has a wave function with macroscopic magnetization, called as the
ferromagnetic phase.

In this paper we focus on the determination of the order of the phase
transition. Supposing first that the possible transitions to the polar
or to the ferromagnetic phases of the spin-1 Bose gas are continuous
phase transitions and restricting ourselves to universal quantities of
the system, the problem can be mapped to a classical field theory with
the following bare action \cite{Landsman1,Andersen1}:
\begin{multline}
  \label{eq:bareact}
  \mathcal{S}[\ph\conj,\ph]\\=\int\d^dx\Big[\frac{1}{2}
  \partial_\mu\ph\conj_r(x)\partial_\mu\ph_r(x)+\frac{1}{2}
  m^2\ph\conj_r(x)\ph_r(x)\\+\frac{c_{rs,r's'}}{4}
  \ph\conj_r(x)\ph\conj_{r'}(x)\ph_{s'}(x)\ph_s(x)
  \Big],
\end{multline}
with $\ph_r(x)$ the 3-component, complex, classical field at the
$d$-dimensional position $x$, and $\partial_\mu$ is the
$d$-dimensional gradient. The dimension of the system is generalized
from $3$ to $d$ for later purposes. Further on we set $\hslash=\kb=1$.
The bare mass of the field theory is denoted by $m^2 \equiv -2 M\mu$,
and the tensor structure of the interaction term takes the form of
\begin{equation}
  \label{eq:int2}
  c_{rs,r's'}=\frac{16\pi^2a_0}{\lb^2}
  \Big(\mathcal{P}_0\Big)^{rs}_{r's'}+
  \frac{16\pi^2a_2}{\lb^2} \Big(\mathcal{P}_2\Big)^{rs}_{r's'},
\end{equation}
with $\lb=\sqrt{2\pi/M T}$ being the de-Broglie wavelength.

The parameters $m^2$ and $c_{rs,r's'}$ of the bare action
\eqref{eq:bareact} are easily obtained from the finite temperature
action \cite{Popov2} corresponding to the Hamiltonian \eqref{eq:ham}
by simply replacing the imaginary time dependent fields with time
independent ones and by integrating out the imaginary time. This
procedure is clearly a restriction to the zero Matsubara frequency
sector of the full field theory. The effects of the higher Matsubara
frequency components would provide a physical cutoff to the field
theory and change the value of the above parameters. These questions
are not discussed in this paper since the order of the phase
transition is not sensitive for the actual value of the above
quantities.

It is more convenient to introduce the matrices $X\equiv\mathcal{P}_0
+ \mathcal{P}_2$ and $Y\equiv\mathcal{P}_2 - 2\mathcal{P}_0$ and to
express the bare interaction \eqref{eq:int2} with the help of them:
\begin{equation}
  \label{eq:int3}
  c_{rs,r's'}=c_n X_{rs,r's'} + c_s Y_{rs,r's'},
\end{equation}
with
\begin{subequations}
  \label{eqs:bareinter}
  \begin{align}
    c_n&=\frac{16\pi^2}{\lb^2} \frac{a_0+2a_2}{3},\\
    c_s&=\frac{16\pi^2}{\lb^2} \frac{a_2-a_0}{3}.
  \end{align}
\end{subequations}
Note that in the case $a_2>a_0$, i.e. in the polar case $c_s>0$, while
in the ferromagnetic case ($a_0>a_2$) $c_s<0$.

\section{Renormalization up to order of two loops}
\label{sec:renorm}

In the following we suppose that on the critical surface, all of the
renormalized masses are zero. This assumption is valid in the absence
of an external magnetic field and when the system is in a paramagnetic
and non-magnetized phase above the (spinor) Bose--Einstein
condensation. The standard renormalization group program of the
massless theory with dimensional regularization and minimal
subtraction \cite{Amit1,Zinn-Justin1} will be carried out for the
classical field theory, described by the bare action
\eqref{eq:bareact} and interaction \eqref{eq:int3}. The free
propagator corresponding to the quadratic part of the bare action then
reads as:
\begin{equation}
  \label{eq:freeprop}
  G_{(0)rs}(p)=\delta_{rs}G\nul(p)=\delta_{rs}\frac{2}{p^2},
\end{equation}
which is just twice the value of the free propagator in a theory with
real fields.

The $n$-point vertex function is denoted by $\Gamma^{(n)}(p_i,c)$,
with $c$ the bare interaction \eqref{eq:int3}, $p_i$ ($i\in\{1\ldots
n\}$) are the wave-numbers of the vertex function. $\Gamma^{(n)}$ has
$n$ spin indices. The spin indices will be omitted most of the time
for notational simplicity. The renormalization conditions for the
vertex functions then read as:
\begin{subequations}
  \label{eqs:renormcond}
  \begin{equation}
    \label{eq:renormcond}
    \Gamma_R^{(n)}(p_i,\kappa,g)=Z^{n/2}(g)\Gamma^{(n)}(p_i,c),
  \end{equation}
  where $Z(g)$ is the field renormalization constant, $\kappa$ is the
  momentum scale, where the renormalization is made, and $g$ stands
  for the dimensionless renormalized coupling constant, having the
  same index structure as the bare one \eqref{eq:int3}. The connection
  between the bare and renormalized coupling constant is established
  by
  \begin{equation}
    \label{eq:bcrc}
    c_{ij,kl}\equiv\kappa^\epsilon \gn_{ij,kl}=\kappa^{\epsilon}G_{ij,kl}(g),
  \end{equation}
\end{subequations}
with $\epsilon=4-d$, $\gn$ is the dimensionless, bare coupling
constant and $G_{ij,kl}(g)$ the coupling constant renormalization
function. The renormalization constants are expanded as power series
of the renormalized coupling constants $g_{ij,kl}$, according to
\begin{subequations}
  \label{eqs:renconsts}
  \begin{align}
    Z(g)&=1+b^{(2)}(g,g),\\
    G_{ij,kl}(g)&=g_{ij,kl}+a^{(1)}_{ij,kl}(g,g)+a^{(2)}_{ij,kl}(g,g,g).
  \end{align}
\end{subequations}
Here we have introduced the following symbolic notation for scalar and
tensorial quadratic quantities:
\begin{subequations}
  \begin{align}
    b(g,g) &= b^{i_1j_1,k_1l_1}_{i_2j_2,k_2l_2} g_{i_1j_1,k_1l_1}
    g_{i_2j_2,k_2l_2},\\
    a^{(1)}_{ij,kl}(g,g)& = \Big(a^{(1)}_{ij,kl}
    \Big)^{i_1j_1,k_1l_1}_{i_2j_2,k_2l_2}
    g_{i_1j_1,k_1l_1} g_{i_2j_2,k_2l_2},
  \end{align}
\end{subequations}
respectively, and similarly for the cubic $a^{(2)}_{ij,kl}(g,g,g)$.
The 2-point and 4-point vertex functions are expanded as:
\begin{subequations}
  \label{eqs:vertfunpert}
  \begin{equation}
    \Gamma^{(2)}_{ij}(p,\gn)=p^2\delta_{ij}\Big[1-\Sigma^{(2)}(\gn,\gn)
    +\Ordo(\gn^3)\Big],\label{eq:vert2p}
  \end{equation}
  \begin{multline}
    \Gamma^{(4)}_{ij,kl}(p_i,\gn)=\kappa^\epsilon\Big[\gn_{ij,kl}+
    d^{(1)}_{ij,kl}(\gn,\gn)\\+d^{(2)}_{ij,kl}(\gn,\gn,\gn)+\Ordo(\gn^4)
    \Big].  \label{eq:vert4p}
  \end{multline}
\end{subequations}
In Eq. \eqref{eq:vert2p} $p^2\c\Sigma^{(2)}(g,g)$ is the two loop
contribution to the self-energy. (The first order term is momentum
independent and cancelled, since we assume that the renormalized
masses are zero.) The constants $d^{(1)}$ and $d^{(2)}$ are the
corresponding 1 and 2 loop contributions to the four-point function.

\begin{figure}[t]
  \centering
  \includegraphics[scale=0.6]{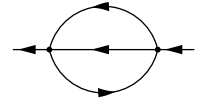}
  \caption{The second order divergent graph contributing to the 2 point function.}
  \label{fig:2pfd}
\end{figure}
\begin{figure}[b]
  \centering
  \includegraphics[scale=0.6]{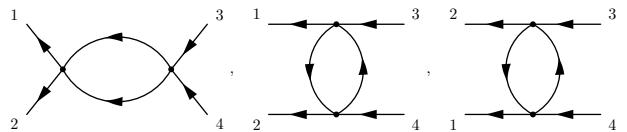}
  \caption{The divergent graphs contributing to the 4 pont function at
    1 loop order}
  \label{fig:4pfd1}
\end{figure}
The divergent graph (up to the order of two loops) contributing to the
two point function is plotted in Fig. \ref{fig:2pfd}. With the
requirement that the renormalized vertex function \eqref{eq:renormcond}
for $n=2$ is finite, one obtains, that
\begin{equation}
  \label{eq:frc2}
  b^{(2)}(g,g)=-\frac{N_d^2}{\epsilon}\frac{g_{im,kl}g_{lk,mi}}{3},
\end{equation}
with $N_d=2/(4\pi)^{d/2}\Gamma(d/2)$ the usual factor appearing after
each momentum integration. $\Gamma(s)$ is the Euler gamma function
with argument $s$.

\begin{figure*}
  \centering
  \includegraphics[scale=0.6,angle=90]{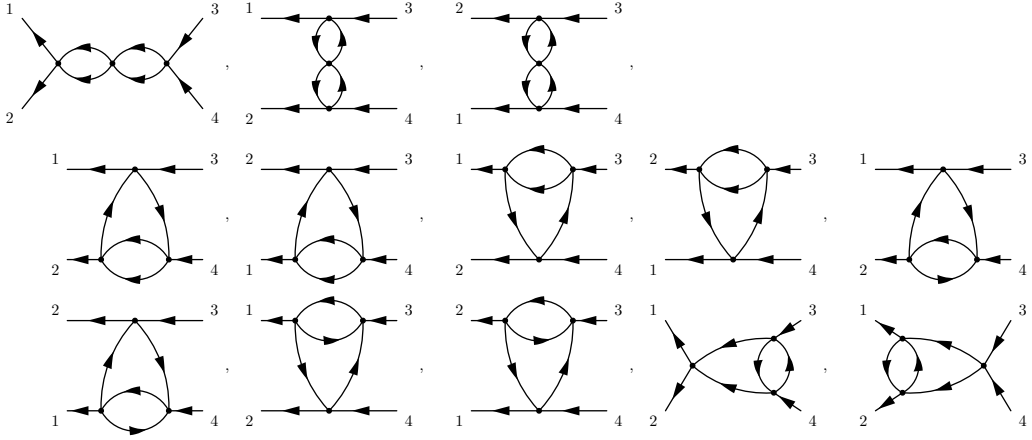}
  \caption{The divergent graphs contributing to the 4 pont function at
    2 loop order}
  \label{fig:4pfd2}
\end{figure*}
The divergent graphs contributing to the four point function at 1 loop
level are drawn in Fig. \ref{fig:4pfd1}, while at 2 loop order are
drawn in Fig. \ref{fig:4pfd2}. With the requirement that the
renormalized four point function [Eq. \eqref{eq:renormcond} with
$n=4$] is finite, the renormalization constants can be obtained.
\begin{subequations}
The one loop contribution is
\begin{equation}
  \label{eq:grenc1}
  a^{(1)}(g,g)=N_d\frac{2}{\epsilon}\Big[g^{(a)}+
  2\big(g^{(b)}+g^{(c)}\big)\Big],
\end{equation}
while the two loop contribution reads as:
  \begin{multline}
    \label{eq:grenc2}
    a^{(2)}(g,g,g)=N_d^2\frac{4}{\epsilon^2}\Big\{g^{(d)}+2\big(g^{(g)}+
    g^{(h)}\big)+g^{(i)}+g^{(j)}\\+g^{(k)}+g^{(l)}+2\big(g^{(m)}+g^{(n)}+
    g^{(o)}+g^{(p)}\big)+8(g^{(e)}+g^{(f)})\\-\epsilon\Big[-g\frac{
      g_{nm,op}g_{po,mn}}{6}+g^{(g)}+g^{(h)}+\frac{1}{2}\big(g^{(i)}+
    g^{(j)}+g^{(k)}\\+g^{(l)}+2\big[g^{(m)}+g^{(n)}+g^{(o)}+g^{(p)}\big]
    \big)\Big]\Big\},
  \end{multline}
\end{subequations}
\begin{subequations}
with the notations
\begin{align}
  g^{(a)}_{ij,kl}&=g_{im,kn}g_{mj,nl},\\
  g^{(b)}_{ij,kl}&=g_{ij,mn}g_{nm,kl},\\
  g^{(c)}_{ij,kl}&=g_{kj,mn}g_{nm,il},\\
  g^{(d)}_{ij,kl}&=g_{im,kn}g_{mo,np}g_{oj,pl},\\
  g^{(e)}_{ij,kl}&=g_{ij,mn}g_{nm,op}g_{po,kl},\\
  g^{(f)}_{ij,kl}&=g_{kj,mn}g_{nm,op}g_{po,il},\\
  g^{(g)}_{ij,kl}&=g_{im,kn}g_{mj,op}g_{po,nl},\\
  g^{(h)}_{ij,kl}&=g_{im,op}g_{kn,po}g_{mj,nl},\\
  g^{(i)}_{ij,kl}&=g_{ij,nm}g_{ko,mp}g_{on,pl},\\
  g^{(j)}_{ij,kl}&=g_{kj,nm}g_{io,mp}g_{on,pl},\\
  g^{(k)}_{ij,kl}&=g_{io,mp}g_{oj,pn}g_{kl,nm},\\
  g^{(l)}_{ij,kl}&=g_{ko,mp}g_{oj,pn}g_{il,nm},\\
  g^{(m)}_{ij,kl}&=g_{ij,nm}g_{kn,po}g_{mp,ol},\\
  g^{(n)}_{ij,kl}&=g_{kj,nm}g_{in,po}g_{mp,ol},\\
  g^{(o)}_{ij,kl}&=g_{io,pm}g_{oj,np}g_{mn,kl},\\
  g^{(p)}_{ij,kl}&=g_{ko,pm}g_{oj,np}g_{mn,il}.
\end{align}
\end{subequations}

\section{Critical properties}
\label{sec:crprop}

The critical properties of the massless theory can be studied with the
help of the $\beta$ function, defined as
\begin{equation}
  \label{eq:betadef}
  \frac{\d G_{ij,kl}}{\d g_{mn,op}}\beta_{mn,op}=-\epsilon G_{ij,kl}.
\end{equation}
The $\beta$ function can be easily expressed by inverting the matrix
$\d G/\d g$ perturbatively in $g$, e.g. with the help of iteration.
The result reads as
\begin{multline}
  \label{eq:beta2}
  \beta=-\epsilon g + 2 N_d \Big[g^{(a)}+2\big(g^{(b)}+g^{(c)}\big)\Big]
  -4 N_d^2 \Big[2\big(g^{(g)}\\+g^{(h)}\big)+g^{(i)}+g^{(j)}+g^{(k)}+
  g^{(l)}+2\big(g^{(m)}+g^{(n)}+g^{(o)}+g^{(p)}\big)\\-g\frac{g_{nm,op}
    g_{po,mn}}{3}\Big]+\Ordo(g^4).
\end{multline}
The tensorial $\beta$ function \eqref{eq:beta2} splits into two
functions, according to
\begin{equation}
  \label{eq:betasplit}
  \beta_{ij,kl}=\beta_n(g_n,g_s)X_{ij,kl}+\beta_s(g_n,g_s)Y_{ij,kl},
\end{equation}
with $X$ and $Y$ defined above Eq. \eqref{eq:int3}. The corresponding
functions are:
\begin{subequations}
  \label{eq:betans}
  \begin{multline}
    \beta_n(g_n,g_s)=-\epsilon g_n+2 N_d \big(7 g_n^2+ 4 g_n g_s + 4
    g_s^2\big)\\-4 N_d^2\big(24 g_n^3 + 22 g_n^2 g_s + 39 g_n g_s^2 +
    20 g_s^3)+ \Ordo(g^4),
  \end{multline}
  and
  \begin{multline}
    \beta_s(g_n,g_s)=-\epsilon g_s+2 N_d \big(6 g_n g_s + 3
    g_s^2\big)\\-4 N_d^2\big(28 g_n^2 g_s + 28 g_n g_s^2 +
    g_s^3)+ \Ordo(g^4).
  \end{multline}
\end{subequations}
These functions are responsible for the flow of the renormalized
coupling constants under a change of scale:
\begin{equation}
  \label{eq:floweq}
  \rho \frac{\d g_{n,s}}{\d \rho}=\beta_{n,s}(g_n,g_s).
\end{equation}
with $\rho$ being the scale, and $g_{n,s}(\rho=1)=g_{n,s}^{0}$. The IR
behavior ($\rho\rightarrow0$) of the system can be studied with the
help of the IR fixed points of the $\beta$ function \eqref{eq:betans}.

\begin{figure}[t!]
  \centering
  \includegraphics{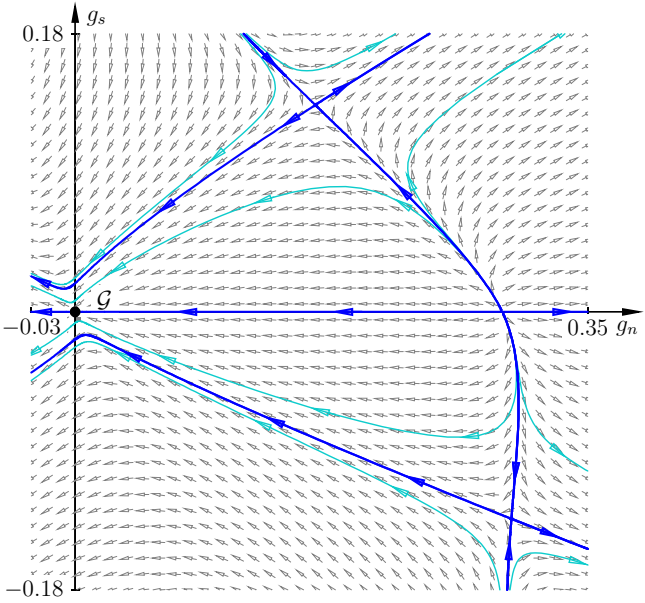}
  \caption{The flow diagram and fixed point structure at $d=4$
    ($\epsilon=0$).}
  \label{fig:flow4d}
\end{figure}

\begin{figure}[t!]
  \centering
  \includegraphics{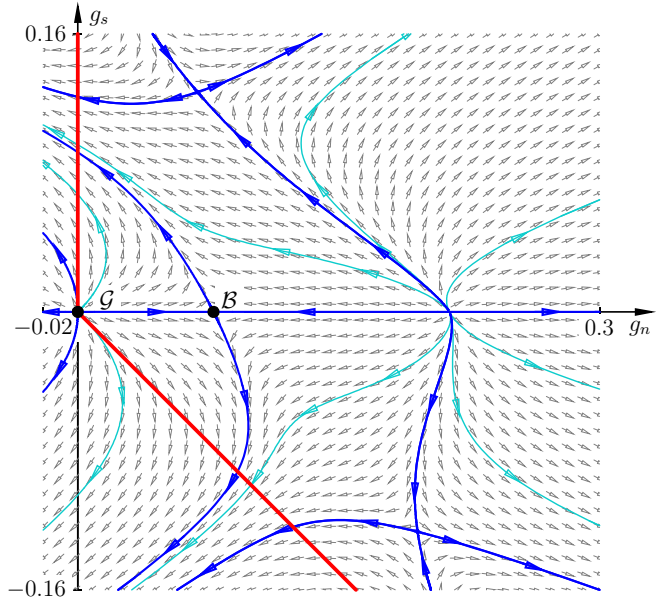}  
  \caption{The flow diagram and fixed point structure at $d<4$ (for
    $\epsilon=0.4$). The red lines indicate the boundary of the
    stability wedge of the zeroth order contribution of the free
    energy (see below).}
  \label{fig:floweps}
\end{figure}

At $d=4$ four real fixed points exist. The Gaussian one:
$\mathcal{G}$, with $(g_n,g_s)=(0,0)$, and three nonphysical fixed
points, which are absent at the one loop calculation. The fixed point
structure and the coupling constant flow is shown in Fig.
\ref{fig:flow4d}. At $d=4-\epsilon$, with $0<\epsilon\ll1$ a new real
fixed point, $\mathcal{B}$, emerges, which is of $\Ordo(\epsilon)$. The
fixed point up to $\Ordo(\epsilon^2)$ is $(g_n,g_s)=(\tilde{g}_n,0)$,
with
\begin{equation}
  \label{eq:u3fp}
  \tilde{g}_n=\frac{1}{2N_d}\epsilon\bigg(\frac{1}{7}+
  \frac{24}{343}\epsilon+\Ordo\Big(\epsilon^2\Big)\bigg).
\end{equation}
This fixed point is the $\U(3)$ symmetric fixed point of Bose
condensation. It is stable from the direction of the Gaussian one,
however it is repulsive through the direction of $g_s$. The fixed
point structure and the flow diagram is shown in Fig.
\ref{fig:floweps}. Since there is no attractive fixed point (for
$g_s\neq0$), the trajectories ``run away'', which is an indication
that both the polar (when $g_s>0$) and the ferromagnetic (when
$g_s<0$) Bose condensation is of first order. Such fluctuation induced
first order transitions are not rare, e.g.  the case of a real
$\phi^4$ theory with cubic anisotropy \cite{Wallace1,KW1}.

The runaway of the trajectories requires an analysis of the free
energy (thermodynamic potential) of the classical system, or as in the
terminology of field theory the effective action, which is the
generating functional of the vertex functions. The free energy is
obtained with the help of the method of steepest descent up to one
loop order \cite{Amit1,Zinn-Justin1}. The zeroth order (tree-graph)
contribution is the bare action \eqref{eq:bareact}:
  \begin{equation}
    \label{eq:ea0loop}
    \frac{1}{V}\Gamma_0[\ph]=\frac{1}{2}m^2 \ph_r\ph_r
    +\frac{1}{4}c_{rs,r's'}\ph_r\ph_s\ph_{r'}\ph_{s'},
  \end{equation}
  with $\ph$ being real and homogeneous. The free energy
  \eqref{eq:ea0loop} describes a continuous phase transition at
  $m^2=0$.  $V=L^d$ is the volume of the system. For $m^2<0$ and
  $c_s>0$ (polar case) the homogeneous field minimizing the potential
  \eqref{eq:ea0loop} can be chosen as $\ph_r = \ph \times (0,1,0)_r$,
  which has zero magnetization, while for $c_s<0$ (ferromagnetic case)
  the minimizing solution can be taken as $\ph_r = \ph
  \times(1,0,0)_r$, which has maximal magnetization. With these
  solutions the free energy at tree-graph level takes the following
  forms:
  \begin{subequations}
    \begin{align}
    \frac{1}{V}\Gamma_0^{\mathrm{pol}}[\ph]&=\frac{1}{2}m^2
    \ph^2+\frac{1}{4}c_n\ph^4,\label{eq:tgepp}\\
     \frac{1}{V}\Gamma_0^{\mathrm{ferr}}[\ph]&=\frac{1}{2}m^2
     \ph^2+\frac{1}{4}(c_n+c_s)\ph^4.\label{eq:tgegpf}
    \end{align}
  \end{subequations}
In the polar case ($c_s>0$) the tree graph thermodynamic potential
\eqref{eq:tgepp} is bounded from below only if $c_n>0$, while in the
ferromagnetic case ($c_s<0$) Eq. \eqref{eq:tgegpf} is confining only
if $c_n+c_s>0$. The line of stability is therefore $c_n=0$ for $c_s>0$
and $c_n+c_s=0$ for $c_s<0$. As can be seen in Fig. \ref{fig:floweps}
all interesting trajectories with $g_s\neq0$ reach the stability
boundaries of the tree graph free energy and go outside the region of
stability. It is therefore mandatory to calculate the next order
contribution.  Following the standard technique
\cite{Amit1,Jackiw1,IA1} (but with complex fields) the one-loop
contribution of the free energy reads as
  \begin{subequations}
    \label{eqs:ea1loop}
    \begin{multline}
      \label{eq:ea1loopp}
      \frac{1}{V}\Gamma_1^{\mathrm{pol}}[\ph]=\frac{1}{2}\int
      \frac{\d^3k}{(2\pi)^3}\bigg\{\log\Big[\big(k^2+m^2+2c_n\ph^2
      \big)^2 \\-
      c_n^2\ph^4\Big]+2\log\Big[\big(k^2+m^2+(c_n+c_s)\ph^2\big)^2 -
      c_s^2\ph^4\Big]\bigg\}
    \end{multline}
    in the polar case, while
    \begin{multline}
      \label{eq:ea1loopf}
      \frac{1}{V}\Gamma_1^{\mathrm{ferr}}[\ph]=\frac{1}{2}\int
      \frac{\d^3k}{(2\pi)^3}\bigg\{\log\Big[\big(k^2+m^2+2(c_n+c_s)
      \ph^2\big)^2 \\-
      (c_n+c_s)^2\ph^4\Big]+2\log\Big[k^2+m^2+(c_n+c_s)\ph^2\Big]\\+
      2\log\Big[k^2+m^2+(c_n-c_s)\ph^2\Big]\bigg\}
    \end{multline}
in the ferromagnetic case.
\end{subequations}
The integrals appearing in Eqs. \eqref{eqs:ea1loop} are divergent.
Carrying out the renormalization scheme renders them finite:
\begin{subequations}
  \begin{equation}
    \label{eq:ea1lr}
    \frac{1}{V}\Gamma_1[\ph]=f(\nu_1)+f(\nu_2)+2f(\nu_3)+2f(\nu_4)
\end{equation}
both for the polar and ferromagnetic cases, with
\begin{align}
  \nu_1^{\mathrm{pol}}&=t+3g_n\ph^2,\\
  \nu_2^{\mathrm{pol}}&=t+g_n\ph^2,\\
  \nu_3^{\mathrm{pol}}&=t+(g_n+2g_s)\ph^2,\\
  \nu_4^{\mathrm{pol}}&=t+g_n\ph^2
\end{align}
in the polar case and
\begin{align}
  \nu_1^{\mathrm{ferr}}&=t+3(g_n+g_s)\ph^2,\\
  \nu_2^{\mathrm{ferr}}&=t+(g_n+g_s)\ph^2,\\
  \nu_3^{\mathrm{ferr}}&=t+(g_n+g_s)\ph^2,\\
  \nu_4^{\mathrm{ferr}}&=t+(g_n-g_s)\ph^2
\end{align}
in the ferromagnetic one. Here we introduced $t$ the renormalized,
dimensionless temperature, and measured the field $\ph$ in units of
$\kappa^{1-\epsilon/2}$, with $\kappa$ being the scale of the
renormalization.
\end{subequations}
The function $f$ appearing in Eq. \eqref{eq:ea1lr} is given by
\begin{equation}
  \label{eq:fdef}
  f(x)=\frac{x^2}{8}\bigg(\log x+\frac{1}{2}\bigg).
\end{equation}
The sum of Eqs. \eqref{eq:ea0loop} and \eqref{eq:ea1lr} gives the free
energy of the classical system \eqref{eq:bareact} in next to leading
order.

On the border of stability the free energy up to one-loop order (both
for the polar and ferromagnetic cases) can be cast to the form (with
neglecting terms not depending on $\ph$):
\begin{equation}
  \label{eq:eprb}
  \frac{1}{V}\Gamma[\ph]=\frac{1}{2}t\ph^2+2f\big(t+2|g_s|\ph^2\big).
\end{equation}
The potential \eqref{eq:eprb} describes systems with a first order
phase transition for $0<|g_s|=\Ordo(\epsilon)$. It is worth to note
that in the Hartree approximation made for the quantum theory of the
spin-1 Bose gas directly in 3 dimensions \cite{KSzSz,SzKSz}, or in the
equivalent lattice mean-field calculation also made directly in 3
dimensions \cite{GK1} the phase transition was found to be of first
order for small coupling constant values (at least for the
ferromagnetic case).

\section{Summary}
\label{sec:sum}

In summary, we have studied the critical properties of spin-1 Bose
gases with the assumption that the phase transition to the polar and
to the ferromagnetic Bose--Einstein condensed phases is of second
order. In this case the universal IR behavior of the quantum system
\eqref{eq:ham} can be studied with the help of a classical field
theory \eqref{eq:bareact} obtained by restricting the fields to the
zero Matsubara frequency sector. The machinery of the
field-theoretical renormalization group then was applied to the
classical field theory. The $\beta$ function was determined up to the
order of two loops in the massless theory. It was found that the only
at least partially stable physical fixed point (for $d<4$) is the
$\U(3)$ symmetric one of Bose--Einstein condensation with $g_s=0$.
However this fixed point was found to be repulsive towards the
direction of $g_s$. This indicates that all trajectories of systems
with nonzero $g_s$ tend towards the border of the classical stability
wedge with successive scale transformations. The free energy [on the
boundary of the stability wedge \eqref{eq:eprb}] of the classical
system \eqref{eq:bareact} was determined in next to leading order in
the method of steepest descent. It was found that both in the polar
and ferromagnetic cases the free energy develops a second local
minimum (besides the trivial one), which turns to be a global minimum
at a certain temperature above $t=0$ (before the second order phase
transition sets in). This shows that the classical system described by
the bare action \eqref{eq:bareact} exhibits a first order phase
transition for $0< |g_s|=\Ordo(\epsilon)$, at least near four
dimensions. This contradicts our assumption that the polar and
ferromagnetic Bose--Einstein condensations are continuous phase
transitions (if they were, the classical description would yield also
a continuous phase transition), and shows that the phase transition is
indeed of first order both for the polar and ferromagnetic gases and
that the jump in the thermodynamic quantities depend on the smaller
coupling constant: $g_s$. One can regard the presented approach as an
indirect proof.

\begin{acknowledgement}
\label{sec:ack}

The present work has been initiated in a visit to the Utrecht
University and was supported by the European Science Foundation under
the project BEC2000+. The author is indebted to prof. H. Stoof for
useful discussions and for his hospitality. The author is also
indebted to profs. A. Patk\'os, L. Sasv\'ari and P. Sz\'epfalusy for
useful discussions and for the careful reading of the manuscript.
Support from the Hungarian National Research Foundation under Grant
No. OTKA T046129 is also kindly acknowledged.
\end{acknowledgement}


\end{document}